\documentstyle[12pt,epsf]{article}
\begin{document}
\def\NCA{\em Nuovo Cimento}
\def\NIM{\em Nucl. Instrum. Methods}
\def\NIMA{{\em Nucl. Instrum. Methods} A}
\def\NPB{{\em Nucl. Phys.} B}
\def\PLB{{\em Phys. Lett.}  B}
\def\PRL{\em Phys. Rev. Lett.}
\def\PRD{{\em Phys. Rev.} D}
\def\ZPC{{\em Z. Phys.} C}
%
\def\st{\scriptstyle}
\def\sst{\scriptscriptstyle}
\def\mco{\multicolumn}
\def\epp{\epsilon^{\prime}}
\def\vep{\varepsilon}
\def\ra{\rightarrow}
\def\ppg{\pi^+\pi^-\gamma}
\def\vp{{\bf p}}
\def\ko{K^0}
\def\kb{\bar{K^0}}
\def\al{\alpha}
\def\ab{\bar{\alpha}}
\def\be{\begin{equation}}
\def\ee{\end{equation}}
\def\bea{\begin{eqnarray}}
\def\eea{\end{eqnarray}}
\def\nn{\nonumber}
\def\slashp{\not \! p}
\def\slashk{\not \! k}
\def\slashq{\not \! q}
\def\slashx{\not \! x}
\def\bi{\begin{itemize}}
\def\ei{\end{itemize}}
\def\CPbar{\hbox{{\rm CP}\hskip-1.80em{/}}}
\hoffset=-1.2cm
\baselineskip=9mm
\hsize=16cm
\vsize=24cm
\begin{titlepage}
\begin{flushright}
June, 2000\\
\end{flushright}
\vskip 1cm
{\centerline{\large{\bf {NON-PERTURBATIVE FERMION PROPAGATOR FOR}}}}
{\centerline{\large{\bf {THE MASSLESS QUENCHED QED3 }}}}
\vskip 1.5cm
\baselineskip=6mm
{\centerline{\large{\bf{A. Bashir }}}}
\vskip 5mm
{\centerline{Instituto de F{\'\i}sica y Matem\'aticas}}
{\centerline{Universidad Michoacana de San Nicol\'as de Hidalgo}}
{\centerline{Apdo. Postal 2-82 , Morelia, Michoac\'an, M\'exico}}
{\centerline{Tel/Fax: +52 4 3271997}}
{\centerline{email: adnan@itzel.ifm.umich.mx}}
\vskip 2cm
{\centerline {ABSTRACT}}
\vskip 3mm
{\leftskip=15mm\rightskip=15mm\noindent

    For massless quenched QED in three dimensions, we evaluate a 
non-perturbative expression for the fermion propagator which agrees
with its two loop perturbative expansion in the weak coupling regime. 
This calculation is carried out by making use of
the Landau-Khalatnikov-Fradkin transformations.
Any improved construction of the 
fermion-boson vertex must make sure that the solution of the 
Schwinger-Dyson equation for the fermion propagator reproduces this
result. For two different gauges, we plot the fermion propagator against 
momentum. We then make a comparison with a similar plot, using the 
earlier expression for the fermion propagator, which takes into account 
only the one loop result. 
\par}
\vskip 1cm
PACS-1996: 11.15.Tk,12.20.-m 
\end{titlepage}

\vfil\eject
\vskip 2cm
\section{Introduction}
\baselineskip=8mm

\indent

    A natural starting point for the non-perturbative study of gauge 
theories is the corresponding set of Schwinger-Dyson equations (SDEs). 
QED in 3-dimensions (QED3) has been a popular choice for such a study 
due to its relative simplicity and its confining behaviour in
the quenched approximation. It  
requires knowledge of the non-perturbative form of the fundamental 
fermion-boson interaction. In the quenched approximation, one can
then calculate the fermion propagator.  Both the propagator and the 
vertex 
must obey essential gauge dependence in accordance with 
Landau-Khalatnikov-Fradkin (LKF)
transformations \cite{LK,Fradkin}. These transformations are written 
in the 
coordinate space 
representation and they allow us to evaluate a non-perturbative expression
for a Greens function in
an arbitrary covariant gauge if we know its value in any particular gauge. 
It seems an insurmountable task to know a Greens function in
any gauge. However, progress can be made in perturbation theory by
calculating it to a certain order.

  Due to the  complicated nature of the LKF
transformations, it has been difficult to derive analytical
conclusions for the vertex. However, the fermion 
propagator 
is relatively easier to analyze and related analytical results 
exist for QED in 3 and 4 dimensions, based upon the knowledge of 
the fermion propagator at the one loop order. These results are 
generally assumed to be true to all orders \cite{BR,Dong}
(referred to as the
{\em transversality condition} in \cite{Burden1}), and
constraints are derived on the non-perturbative form of the
fermion-boson vertex. Realizing that this condition would not 
hold to all orders, Bashir {\em et. al.} have derived
constraints on the vertex in QED4 by demanding general constraints
from the multiplicative renormalizability of the fermion propagator 
\cite{BKP}. Recently, a two loop calculation has been 
done for the fermion propagator in the massless quenched QED3 \cite{AB3,AB4}, 
showing explicitly that the transversality condition is violated. Exploiting
this calculation, we go beyond one loop and present the evaluation of 
the non-perturbative
fermion propagator through the use of LKF transformations. 
In comparison with the corresponding expression in \cite{BR},
our result has an added piece which makes sure that in the weak
coupling regime, correct two loop behaviour of the fermion
propagator is achieved. Our calculation calls for the need to 
construct an improved  vertex which would reproduce
our result when used in the corresponding SDE.
We also plot these two
expressions as a function of fermion momentum and compare the results
for two different gauges.
  
\section{Fermion-Boson Vertex and Gauge Invariance}

      The study of the fermion propagator in quenched QED requires
making an {\em ansatz} for the vertex. An acceptable {\em ansatz} must 
ensure the inclusion of the key features required of it. 
We shall only focus on the features relevant to the discussion in 
this paper:

\bi

\item

    The vertex $\Gamma^{\mu}(k,p)$ must satisfy the 
Ward-Green-Takahashi 
Identity (WGTI) which relates it to the fermion propagator 
$S_F(p^2)$:
\bea
    q_{\mu} \Gamma^{\mu}(k,p) &=& S_F^{-1}(k) - S_F^{-1}(k) \;,
\eea
where ${q=k-p}$.

\item 

   It must reduce to the Feynman expansion of the perturbative vertex in
the weak coupling regime.

\item

   It must ensure local gauge covariance of the propagators and vertices.

\ei
Although the WGTI is a consequence of gauge invariance, 
it only fixes the longitudinal part $\Gamma_L(k,p)$ of the complete vertex
\cite{BC}, $
\Gamma^{\mu}(k,p)=\Gamma^{\mu}_{L}(k,p)+\Gamma^{\mu}_{T}(k,p) $,
whereas the transverse part $\Gamma_T(k,p)$, defined by the equation
$ q_{\mu}\Gamma^{\mu}_{T}(k,p)=0$,
remains undetermined. Without a proper choice of this part, one cannot 
ensure the local gauge covariance of the propagators and the vertex
as demand the LKF transformations. Unfortunately, the LKF transformation
law for the vertex is too complicated to be made use of. However, 
the corresponding rule for the fermion propagator is relatively simple.
A proper choice of the transverse vertex in an arbitrary gauge 
is essential to satisfy it, as it is related to the fermion propagator 
through the following SDE in the Euclidean space,
Fig.~(1):
\begin{eqnarray}
  S_F^{-1}(p) &=&  {S_F^0}^{-1}(p) + e^2 \; \int \frac{d^3k}{(2 \pi)^3}
  \; \Gamma^{\mu}(k,p) \,  S_F(k) \, \gamma^{\nu} \, \Delta_{\mu \nu}^0(q)
  \qquad,
\end{eqnarray}
where $S_F^0(p)= 1/{i \slashp}$ and we express $
 S_F(p)=F(p^2)/i \slashp.$
The photon propagator can be split into the transverse and the
longitudinal parts as:
\begin{eqnarray}
     \Delta_{\mu \nu}^0(q) &=&  {\Delta_{\mu \nu}^0}^T(q) + \xi \; 
     \frac{q_{\mu}q_{\nu}}{q^4} \qquad,
\end{eqnarray}
where $
      {\Delta_{\mu \nu}^0}^T(q) =  \left[
 \delta_{\mu\nu}- q_{\mu}q_{\nu}/q^2\right] / q^2 .$
Burden and Roberts \cite{BR} pointed out that the condition
\begin{eqnarray}
\int \frac{d^3k}{(2 \pi)^3}
  \; \Gamma^{\mu}(k,p) \,  S_F(k) \, \gamma^{\nu} \, 
  {\Delta_{\mu \nu}^0}^T(q)&=&0 \;,
\end{eqnarray}
the so called transversality condition  \cite{Burden1}, leads to the 
correct LKF behaviour of the fermion propagator. This condition ensures
that $F(p^2)=1$ in the Landau gauge. The LKF transformations then yield
\begin{eqnarray}
F(p^2)&=&1-\frac{\alpha\xi}{2 p} \; 
{\rm tan}^{-1} \left[ \frac{2 p}{\alpha\xi} \right] \qquad.
\end{eqnarray}
However, it has been shown in \cite{AB3,AB4} that although this 
condition is satisfied at one loop order, it gets violated at the 
two loop order, leading to the following expression for the 
fermion propagator \footnote{An error made in the first reference of 
\cite{AB3} was 
corrected in the second and in \cite{AB4}}:
\begin{eqnarray}
   F(p^2)&=& 1 - \frac{\pi \, \alpha \xi}{4 p}  + 
               \frac{\alpha^2 \xi^2}{4p^2}
              - \frac{3 \alpha^2}{4 p^2} \left(\frac{7}{3}-\frac{\pi^2}{4} 
              \right) +{\cal O}(\alpha^3) \,.
\end{eqnarray}
This expression for the fermion propagator of course does not satisfy the 
LKF transformations in a non-perturbative fashion. However, making use of the said
transformations, we can calculate an expression for
the fermion propagator which does transform non-perturbatively as require
the LKF transformations, which is an important requirement of gauge covariance.
We carry out this exercise in the next section.

\section{Propagator and the LKF transformation}

Let us use the following notation and definition of the  massless fermion 
propagator in the momentum and
coordinate spaces respectively, in an arbitrary covariant gauge $\xi$:
\bea
\nn  S_F(p;\xi) &=& \frac{F(p;\xi)}{i \slashp} \;,  \\ 
     S_F(x;\xi) &=& \slashx X(x;\xi) \;.
\eea
These expressions are related by the following Fourier transforms:
\bea
\nn     S_F(p;\xi) &=& \int d^3x \; {\rm e}^{i p \cdot x} \; S_F(x;\xi) \\
        S_F(x;\xi) &=& \int \frac{d^3p}{(2 \pi)^3} \; {\rm e}^{-i p \cdot x} 
                       \;  S_F(p;\xi) \quad.
\eea
The LKF transformation relating the coordinate space fermion propagator
in the Landau gauge to the 
coordinate space fermion propagator
 in an arbitrary covariant gauge reads:
\bea
  S_F(x;\xi)&=&  S_F(x;0) \; {\rm e}^{-(\alpha \xi/a)x} \qquad.
\eea
 The following are the steps to find the non-perturbative expression for
the fermion propagator in momentum space in an arbitrary covariant gauge:
(i) Input the perturbative expression for the fermion propagator in
the Landau gauge, i.e., $F(p;0)$.
(ii) Evaluate $X(x;0)$ by taking the Fourier transform.
(iii) Calculate $X(x:\xi)$ by using the LKF transformation law.
(iv) Fourier transform back the result to $F(p:\xi)$.
\noindent
Eq.~(6) implies that
\bea
    F(p;0) &=& a_0 \; + \; a_1 \, \frac{\alpha}{p} \; + \; a_2 \, 
\frac{\alpha^2}{p^2} \; + \; {\cal O}(\alpha^3) \;,
\eea
where $a_0 = 1,  a_1 = 0 $ and $a_2=-7/4+3 \pi^2/16$.
Although $a_1=0$, we shall keep this term in order to prove a point later.
Eq.~(8) permits us to carry out the Fourier transform of Eq.~(10). On doing that and
carrying out the angular integration, we get
\bea
\nn  X(x;0)  
&=& - \sum_{n=0}^{n=2} \; \frac{a_n \alpha^n}{2 \pi^2 x^3} \; 
\int_0^{\infty} \;
 \frac{dp}{p^{n+1}}  \;
\left(\, {\rm sin}px \, - \, px \, {\rm cos}px \, \right) \;.
\eea
The radial integration then yields:
\bea
   X(x;0) &=&  - \frac{a_0}{4 \pi} \, \frac{1}{x^3} \; - \;  \frac{a_1}{8 \pi}\, 
\frac{\alpha}{x^2} \; - \; \frac{a_2}{8 \pi} \, \frac{\alpha^2}{x} \;.
\eea
This result in the Landau gauge is related to that in arbitrary covariant
gauge through the LKF transformation, Eq.~(9).
In order to Fourier transform the result back to the momentum space,
we use Eq.~(8). Substitute Eq.~(9) in it, multiply the equation by 
$\slashp$ and then
take the trace:
\bea
    F(p;\xi) &=& i \int d^3x \;  p \cdot x \;  {\rm e}^{i p \cdot x} \; X(x;0)
  \;  {\rm e}^{-(\alpha \xi/2)x} \quad.
\eea
Substituting Eq.~(11) in it and carrying out the integrations, we obtain:
\bea
 F(p;\xi) &=& a_0\;  - \; \frac{\alpha (\pi \xi -4 a_1) }{2 \pi p} \,  
\,{\rm tan}^{-1} \, 
\frac{2p}{\alpha \xi} 
\;-\;  \frac{4 \alpha^2}{\alpha^2 \xi^2 + 4p^2} \, \left[ \frac{a_1 \xi}{\pi}
 -  \frac{4a_2 p^2}{\alpha^2 \xi^2 + 4p^2} 
\right] \;.
\eea
Expanding this expression around small values of $\alpha$ we get
\bea
    F(p, \xi) &=& a_0 + a_1 \frac{\alpha}{p} -  \frac{\pi \xi \alpha}{4p}
 -a_1 \frac{\alpha^2 \xi}{\pi p^2} + a_2 \frac{\alpha^2}{p^2} + 
  \frac{\alpha^2 \xi^2}{4 p^2} \;+ \; {\cal O}(\alpha^4) \;.
\eea
For $\xi=0$, we recuperate Eq.~(10). There are some interesting points to note:
\bi

\item

The constant $a_1$ appears as a coefficient of ${\cal O}(\alpha)$ term
as well as ${\cal O}(\alpha^2 \xi)$ term in Eq.~(14). The fact that $a_1=0$ 
 thus automatically rules out the presence of 
${\cal O}(\alpha^2 \xi)$ term.

\item

Apart from the $a_1$ terms, $\xi$ always appears in conjunction with
$\alpha$, i.e., in the form $\alpha \xi$.

\item 

In the perturbative expression, Eq.~(14), there exists no $\alpha^3$ term,
and the same is true for higher odd powers of $\alpha$. This does not
of course rule out the possibility of encountering these terms on perturbative
evaluation of $F(p,\xi)$ at the three-loop level, and so on.

\ei
 Substituting the values of $a_0$, $a_1$ and $a_2$,
we arrive at the following final result:
\bea
 F(p;\xi) &=& 1\;  - \; \frac{\alpha \xi}{2p} \, {\rm tan}^{-1} \, 
\frac{2p}{\alpha \xi} \;-\;  
\frac{(28-3 \pi^2)p^2 \alpha^2}{(\alpha^2 \xi^2 + 4p^2)^2} \;.
\eea
This expression contains an expected additional term in comparison with
Eq.~(5).  This term ensures that the perturbative expansion of Eq.~(15)
matches correctly on to the two-loop calculation of $F(p;\xi)$. 
It also has the correct gauge dependence non-perturbatively as 
demanded by LKF transformations, in contrast with Eq.~(6). Moreover, being
non-perturbative in nature, it contains exact information of terms
of orders higher than $\alpha^2$. For example, the
perturbative expansion of Eq.~(15) to ${\cal O}(\alpha^4)$ reads:
\begin{eqnarray}
   F(p^2)&=& 1 - \frac{\pi \, \alpha \xi}{4 p}  + 
               \frac{\alpha^2 \xi^2}{4p^2}
              - \frac{3 \alpha^2}{4 p^2} \left(\frac{7}{3}-\frac{\pi^2}{4} 
              \right) - \;   \frac{\alpha^4 \xi^4}{48 p^4}  
               \nonumber \\ \nonumber \\
  &-& \frac{3 \alpha^4 \xi^2}{8p^4} \left( \frac{7}{3} - \frac{\pi^2}{4}
 \right) + {\cal O}(\alpha^6)  \,.
\end{eqnarray}
Note that there are no $\alpha^3$ terms in this expression. Therefore, in
conjunction with the structure of LKF transformation, we conclude that
in the actual perturbative calculation of $F(p^2)$ to ${\cal O}(\alpha^3$),
there will be no terms of the type $\alpha^3 \xi^3$, $\alpha^3 \xi^2$,
or $\alpha^3 \xi$. This information is not contained in Eq.~(6), which 
of course does not know anything about orders higher than 
$\alpha^2$. Similarly for ${\cal O}(\alpha^4)$, Eq.~(15) exactly gives the
coefficients of $\alpha^4 \xi^4$, $\alpha^4 \xi^3$ and
$\alpha^4 \xi^2$ terms in perturbation theory, and so on for higher
order terms. 

	In perturbation theory, every higher order term
is expected to be much smaller than the term in the previous order
in a systematic way. 
Naturally, one wonders what is the relative contribution of 
the additional piece in Eq.~(15) in the non-perturbative regime.
In Figs.~(2,3), we have drawn $F(p^2)$ as obtained from Eq.~(5)
and Eq.~(15) for two different values of the gauge parameter. For larger
values of the gauge parameter, the two results start merging into each 
other. However, for low values of the gauge parameter, a bump arises
in the $F(p^2)$ given by Eq.~(15) at low values of $p$ because of the 
maximum in the additional piece at $p=\alpha \xi/2$. Therefore,
one concludes that for higher values of the gauge parameters, the
additional piece modifies  Eq.(5)  insignificantly.
However, for decreasing values of the gauge parameter, the difference 
starts increasing for low momenta. In essence,  Figs.~(2,3)
display the values of the gauge parameter for which the more complete
Eq.~(15) will deviate significantly from Eq.~(5) in the non-perturbative
regime.

\section{Conclusions}

    In this paper, we present the calculation of the non-perturbative
fermion propagator using the knowledge of its two-loop expansion and
the LKF transformations.  It is natural to assume that physically meaningful 
solutions
of the Schwinger-Dyson equations must agree with perturbative results
in the weak coupling regime. This realization has been used in
QED4 \cite{BKP,CP,KRP}, and more recently in QED3 \cite{AB3,AB4}, to
use perturbation theory as a guide towards the non-perturbation truncation 
of Shwinger-Dyson equations. So far, progress has been made in this 
context by attempting to
make the correspondence of non-perturbative propagators and vertex
to their one-loop expansion. In this paper, we have gone beyond the 
one-loop order and constructed a fermion propagator which agrees with
perturbation theory at least up to two-loops, and also has the correct
gauge dependence as demanded by its LKF transformations. On the numerical 
side, it is important to know the contribution of the new piece in 
Eq.~(15) as compared to the rest of the equation. It turns out that for
higher values of the gauge parameter we do not need to worry about it.
However in the neighbourhood of the Landau gauge, it causes significant 
deviation of the total result in comparison with the one in its absence
for low values of momentum.

    As the fermion propagator is related to the 3-point vertex through
its SDE, our results put constraints on the possible
forms for the unknown transverse part of the vertex. This part is in 
principle determined by understanding how the essential 
gauge dependence of the vertex demanded by its LKF
transformation is satisfied non-perturbatively. In practice it is
not an easy condition to implement. However, a simpler constraint is that
any non-perturbative construction of the transverse vertex must ensure
that we recuperate Eq.~(15) when used in the SDE  for the fermion 
propagator, leading to a more reliable  non-perturbative truncation of
SDEs.

\vspace{5mm}
\noindent
{\large{\bf{Acknowledgements:}}}
I would like to acknowledge the CIC and CONACYT grants under the 
project 32395-E.

\newpage

{\centerline{\LARGE{\bf{Figure Captions}}}}
\vspace{5mm}

\begin{enumerate}
\item  Schwinger-Dyson equation for fermion propagator in 
quenched QED. 
\item  $F(p:\xi)$ as a function of $p$ for $\xi=1$. The dotted 
and the continuous lines correspond to Eq.~(5) and Eq.~(15) respectively.
\item $F(p:\xi)$ as a function of $p$ for $\xi=0.5$. The dotted 
and the continuous lines correspond to Eq.~(5) and Eq.~(15) respectively.
\end{enumerate}

\vspace{2.5cm}

{\centerline{\LARGE{\bf{ Figures}}}}


\vspace{2.5cm}
\noindent
{\centerline{Fig. 1}}

\epsfbox[90 50 -125 240]{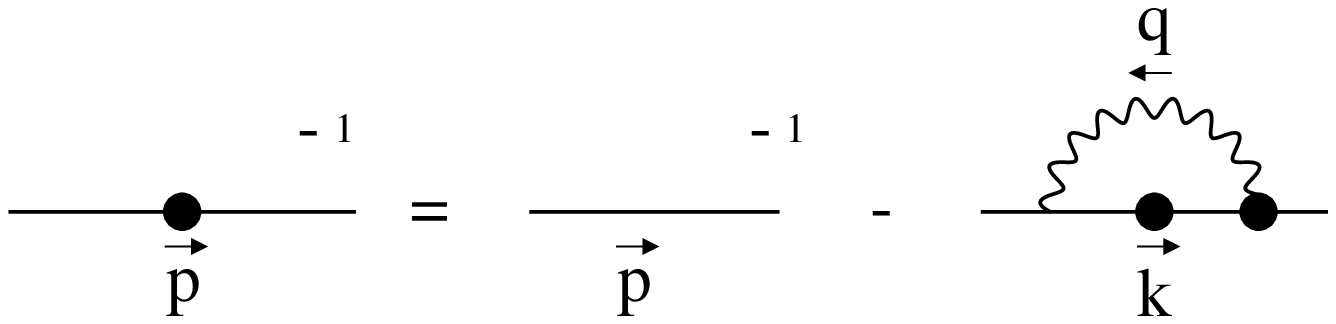}
%

\epsfbox[90 50 -125 720]{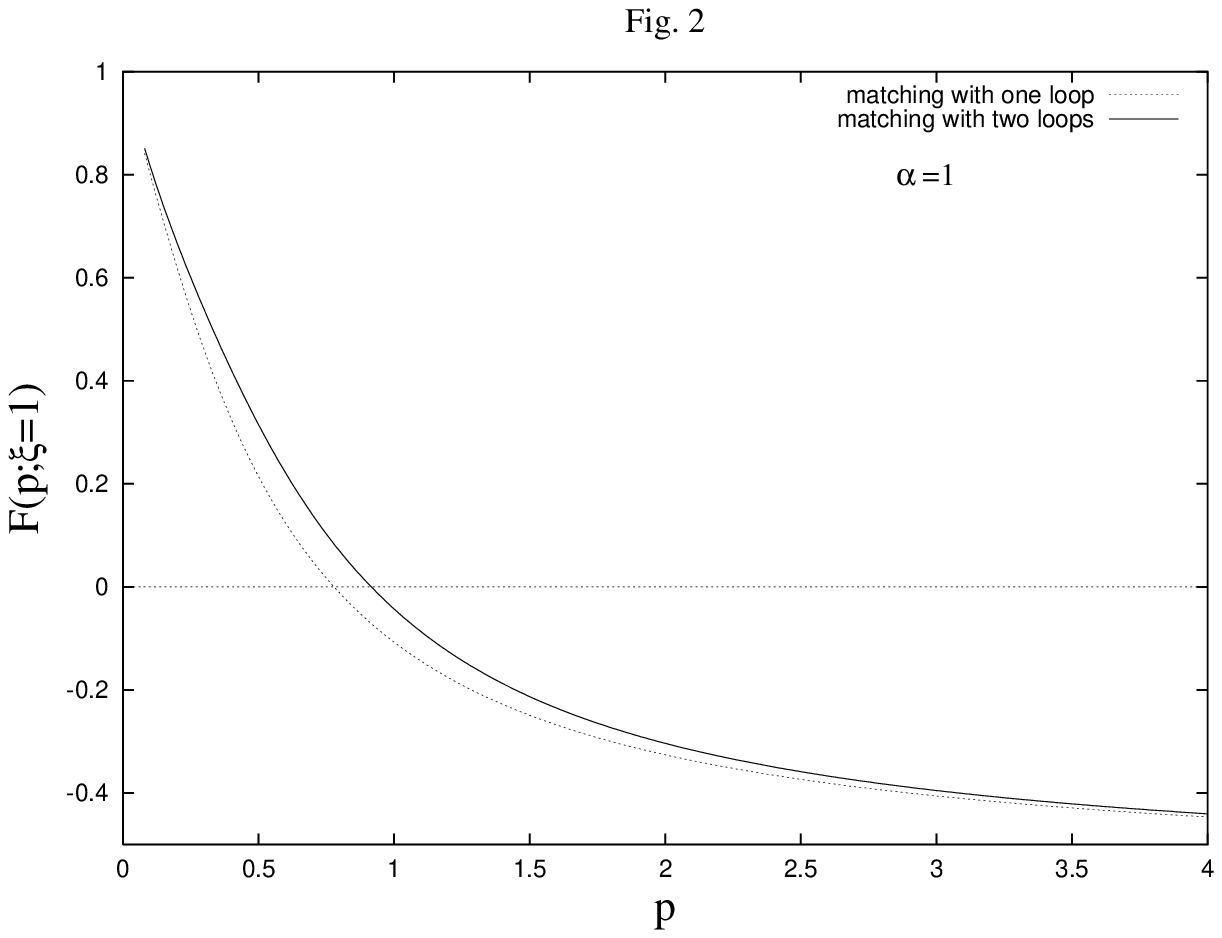} 
\epsfbox[90 400 -125 720]{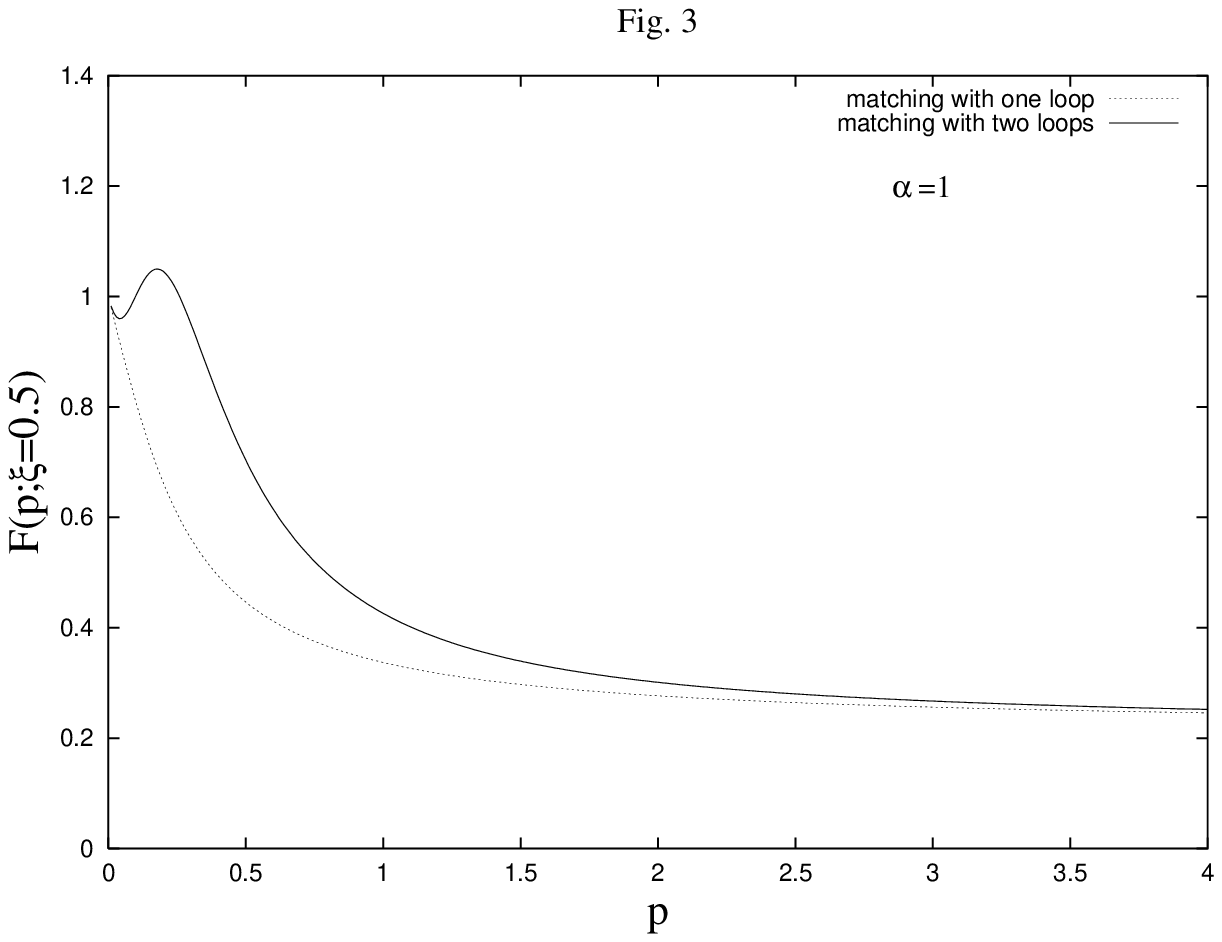}


\begin{thebibliography}{55}
\bibitem{LK} L.D. Landau and I.M. Khalatnikov, Zh. Eksp. Teor. Fiz. {\bf 29},
89 (1956)
[Sov. Phys. JETP, {\bf 2}, 69 (1956)],\\
B. Zumino, J. Math. Phys. {\bf 1}, 1 (1960).
\bibitem{Fradkin} E.S. Fradkin, Sov. Phys. JETP, {\bf 2}, 361 (1956).
\bibitem{BR} C.J. Burden and C.D. Roberts, Phys. Rev. {\bf D47}, 5581 (1993).
\bibitem{Dong} Z. Dong, H.J. Munczek and C.D. Roberts, Phys. Lett. 
{\bf B333}, 536 (1994).
\bibitem{Burden1} C.J. Burden and P.C. Tjiang, 
Phys. Rev. {\bf D58}, 085019 (1998).
\bibitem{BKP} A. Bashir, A. K{\i}z{\i}lers{\"u}
and M.R. Pennington, Phys. Rev. {\bf D57}, 1242 (1998). 
\bibitem{AB3} A. Bashir, A. K{\i}z{\i}lers\"{u} and M.R. Pennington, 
ADP-99-8/T353, DTP-99/76, hep-ph/9907418; ADP-00-29/T412, DTP-00/30. 
\bibitem{AB4} A. Bashir, to appear in the 
Proceedings of
the Workshop on Light-Cone QCD 
and Non-perturbative Hadron
Physics, University of Adelaide, Adelaide, Australia (1999).
\bibitem{BC} J.S. Ball and T.-W. Chiu, Phys. Rev. {\bf D22}, 2542 (1980).
\bibitem{CP} D.C. Curtis and M.R. Pennington, Phys. Rev. {\bf D42}, 4165 (1990).
\bibitem{KRP} A. K{\i}z{\i}lers\"u, M. Reenders and M.R. Pennington, Phys.
Rev. {\bf D52}, 1242 (1995).
%
\end{thebibliography}
\end{document}